\documentclass[prb,amssymb,twocolumn,superscriptaddress]{revtex4}
\usepackage{amsmath}
\usepackage{amssymb}
\usepackage{subfigure}
\usepackage{graphicx}
\usepackage{xcolor}

\newcommand{\be}{\begin{equation}}
\newcommand{\ee}{\end{equation}}

\newcommand{\lf}{\left}
\newcommand{\rg}{\right}
\newcommand{\ra}{\rangle}
\newcommand{\la}{\langle}

\newcommand{\bea}{\begin{eqnarray}}
\newcommand{\eea}{\end{eqnarray}}

\newcommand{\nn}{\nonumber}

\renewcommand{\l}{\lambda}

\begin{document}

\title{Rashba-metal to Mott-insulator transition}

\author{Valentina Brosco}
\affiliation{Istituto dei Sistemi Complessi (ISC-CNR), Via dei Taurini 19, I-00185 Roma, Italy}
\affiliation{Scuola Internazionale Superiore di Studi Avanzati (SISSA) and 
CNR-IOM DEMOCRITOS, Istituto Officina dei Materiali, Consiglio Nazionale delle Ricerche,
Via Bonomea 265, 34136 Trieste, Italy}

\author{Massimo Capone}
\affiliation{Scuola Internazionale Superiore di Studi Avanzati (SISSA) and 
CNR-IOM DEMOCRITOS, Istituto Officina dei Materiali, Consiglio Nazionale delle Ricerche,
Via Bonomea 265, 34136 Trieste, Italy}

\date{\today}


\begin{abstract} 
The  recent discovery of materials  featuring strong Rashba spin-orbit coupling (RSOC) and strong electronic correlation raises questions about the interplay of Mott and Rashba physics. 
In this work, we employ cluster perturbation theory to investigate  the spectral properties of the two-dimensional Hubbard model in the presence of a significant or large RSOC. We show that RSOC strongly favors metallic phases and competes with Mott localization, leading to an unconventional scenario for the Mott transition which is no longer controlled by the ratio between the Hubbard $U$ and an effective bandwidth. The results show a strong sensitivity to the value of the RSOC.
\end{abstract}

\maketitle

\section{Introduction}
The breaking of inversion symmetry has important consequences on the properties of matter. 
%
%
%
Just to mention few examples, it can lead to unconventional superconducting pairing, \cite{gorkov2001,fischer2018} it controls magnetic ordering at interfaces and surfaces,\cite{hellman2017} it rules the generation of spin currents, \cite{beenakker1997,brosco2010,gorini2017} and it determines the locking of spin and quasi-momentum in metals with strong Rashba coupling.\cite{manchon2015,WinklerSpinOrbitCoupling}
Furthermore, inversion symmetry breaking effects can be controlled and enhanced  by  material engineering \cite{ast2007,liu2013,tresca2018,rashba2012, generalov2017,ishizaka2011}  and gating.\cite{nitta1997,caviglia2010}
%
In a large class of materials and heterostructures,\cite{generalov2017,santoscottin2016,tresca2018, ishizaka2011} inversion symmetry breaking and its effects coexist with electron-electron correlation. This calls for a systematic study of the interplay between these two effects, which, on one hand, can help us to understand parity-violating phenomena in actual solids where interactions are significant and on the other, owing to the intrinsic tendency of correlated systems towards magnetic ordering, holds a huge potential for the development of antiferromagnetic spintronics.\cite{baltz2018,jungwirth2018}\\
Motivated by these findings, in the present work we focus on a well-known consequence of inversion symmetry breaking, namely, Rashba spin-orbit coupling (RSOC), \cite{rashba1959} and we show that it significantly affects the physics of the metallic and Mott insulating phases and, consequently of the Mott transition connecting the Rashba metal and the Mott insulator. In order to highlight the intrinsic correlation effects, we will restrict to paramagnetic solutions without magnetic ordering.

Previous works describing the interplay of RSOC and electronic correlation were mainly focused on the magnetic phase diagram,\cite{xinzhang2015,farrell2014} on the investigation of topological effects \cite{laubach2014,rosenberg2017} and  on the  properties of the associated Fermi liquid. \cite{ashrafi2012,ashrafi2013}
Ref.[\onlinecite{xinzhang2015}] demonstrates that RSOC favors the onset of a metallic phase at weak Hubbard interaction and it modifies the magnetic structure of the insulating phase, in qualitative agreement with a static mean field approach. \cite{farrell2014}

  
As a simple approach which allows to study the effects of  strong on-site interaction  without spoiling the non-abelian gauge structure induced by the SOC, we employ cluster perturbation theory (CPT).  The low-numerical cost of this method allows us to scan  a wide  region of parameters, ranging from the weakly correlated Rashba metal  to the Rashba-Mott insulator phase.   
We thus investigate  two complementary aspects of the interplay of Rashba SOC and electronic correlation.
In the Mott insulator phase,  we show that, due to the breaking of SU(2) spin symmetry, Rashba SOC yields a mixing of singlet and triplet  resonating valence bond (RVB)  states possibly opening a new screening channel of local interactions related to the {\sl Pauli screening}  discussed in Ref. [\onlinecite{brosco2018}]. 
At large Rashba coupling and strong interaction where the system realizes a correlated metallic phase,  we instead demonstrate that the breaking of parity associated with Rashba  SOC enables two kinds of low-energy fermionic excitations and it results in a pseudogap phase.\\
%
%
%
%
%
In both metallic and insulating phases we find that RSOC counteracts localization effects,  yielding qualitative and quantitative modifications in the Mott transition. 
Our work thus hints at an enhancement of transport in strongly correlated systems in the presence of RSOC, opposite to what happens in weakly interacting disordered Fermi gases. \cite{brosco2016,brosco2019,brosco2017a} It can be therefore extremely relevant to account for the transport properties of  oxides heterostructures, \cite{pai2017} surface alloys \cite{tresca2018} and polar semiconductors. \cite{santoscottin2016,ishizaka2011,maass2016} Furthermore it suggests  the possibility of exploiting the tunability of Rashba spin-orbit coupling to control transport in strongly correlated materials.

The paper is organized as follows. After describing the model and the method in Section \ref{sec-model}, in Section \ref{sec-results} we present our main results concerning the structure of the spectrum and the density of states across the metal-insulator transition. We then discuss the peculiarities of the  Rashba-Mott insulator in Section \ref{sec-ins}  while  in section \ref{sec-met} we focus on the correlated metallic phase.
Eventually in the Appendices we discuss technical details on the method.

 \begin{figure}[t!]
\begin{center}
\includegraphics[width=0.35\textwidth]{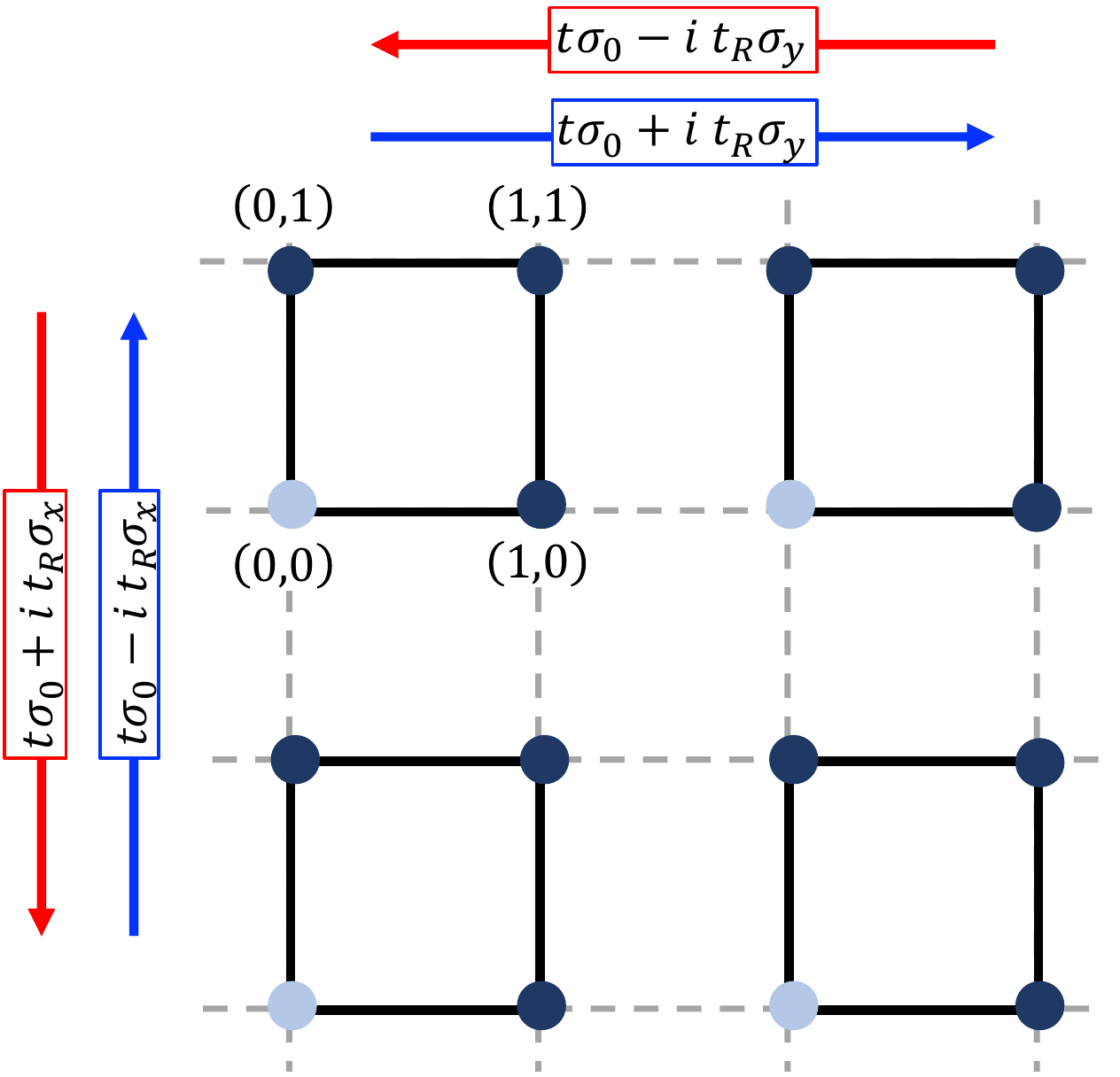}
\caption{Plaquette tiling  of the square lattic. Both intra and inter plaquette tunneling include a standard spin-diagonal and a spin dependent tunneling. Inter-plaquette tunneling is treated perturbatively.}\label{fig-tiling}
\end{center}
\end{figure} 
%
\section{Model}
\label{sec-model}

\noindent We consider the following Rashba-Hubbard model 
\be\label{eq:ham}
H=H_0+H_U.
\ee
where $H_U$ is the Hubbard interaction,
\be H_U=U\sum_{i}n_{i\uparrow}n_{i\downarrow}, \ee %
while $H_0$ can be written as  the sum of a  standard spin-diagonal  hopping and a  spin-flip Rashba hopping term between nearest-neighboring sites as follows
\be
\label{hopping+Rashba}
H_{0}=-t\sum_{\la ij\ra}c^\dag_i c_j-t_R\sum_{\la ij\ra}c^\dag_i (\vec \alpha_{ij} \times \vec \sigma)_{z}c_j
\ee
 where we introduced the local spinor creation and annihilation operators $c^\dag_i$ and $c_i$ and we defined the vector $\vec \alpha_{ij}=(\alpha^x_{ij},\alpha^y_{ij})$ with
$\alpha_{ij}^\mu=i(\delta_{ij+a_\mu}-\delta_{ij-a_\mu})$ where $a_\mu$ is the unitary translation in the $\mu$ direction.
 
 We use cluster perturbation theory \cite{senechal2000,senechal2002} (CPT) as a simple and computationally extremely cheap method which is able to capture the competition between the inherently non-local physics described by $H_0$ and the the local interaction $H_U$ which drives the system towards Mott localization.

Within this approach the lattice is partitioned into a superlattice of identical clusters and  the Green's function $G(\omega,{\bf k})$ is computed solving exactly the cluster and,  treating the intercluster hopping perturbatively. 
Specifically, we  use  four-site plaquette clusters that, as stressed in Ref.[\onlinecite{brosco2018}],  are the minimal clusters where the non-abelian gauge structure of Rashba coupling can emerge. With this choice,  our study encompasses the Pauli screening mechanism discussed in  Ref.[\onlinecite{brosco2018}]  and it preserves the basic symmetries of the lattice.

We  tile the lattice as shown in Fig. \ref{fig-tiling}.  
and 
we decompose the lattice vectors as 
${\bf r}_i={\bf r}_{m}+ {\bf r}_{a} $
where ${\bf r}_{m}$ enumerates the plaquette superlattice sites (light-blue dots  in Fig. \ref{fig-tiling}) and it refers to the position of the lowermost left site of each plaquette  while ${\bf r}_{a}$ indicates the position of the site in the plaquette.
The Hamiltonian is thus partitioned as follows:
\be
H=H_{\rm loc}+V \label{partition}
\ee
where $H_{\rm loc}$ contains all the intra-cluster terms (diagonal in the index $m$) including interaction while $V$ accounts for the interplaquette hopping.  
The matrix $V$ and the local Hamiltonian are defined to guarantee current conservation upon tunneling along $x$ and $y$. 
In particular, as schematically indicated by the blue and red arrows in Fig. \ref{fig-tiling} tunneling in opposite directions yields opposite spin rotations. More details on the partitioning are given in Appendix \ref{app-A}.

Following the route suggested {\sl e.g.}  in Ref. [\onlinecite{senechal2002}],   we  perform  a partial Fourier transformation with respect to the cluster position indices describing the Hamiltonian $H$ in the mixed representation.\\  
In this representation  $V$ can be recast as follows
\be
V=-\!\sum_{{\bf k},a,b}c^\dag_{a{\bf k}}\tilde T_{ab}({\bf k})c_{b{\bf k}}.
\ee
where ${\bf k}$ belongs to the Brillouin zone of the original lattice while the interplaquette hopping amplitude $\tilde T_{ab}({\bf k})$  is represented by a $2\times2$  matrix spin space.
The full interplaquette hopping matrix  has thus dimension 8 and it can be written as follows
\bea
\tilde T({\bf k})&=&\lf(e^{-2 ik_x}(t \sigma_0+it_R\sigma_y)\otimes \tau_x\rg.\nonumber\\&&\lf.+e^{-2 ik_y}(t \sigma_0-it_R\sigma_x)\otimes\tau_y+{\rm H.c.}\rg)\label{T(k)}\eea
with the matrices $\tau_x$ and $\tau_y$ denoting forward unitary translations in the $x$ and $y$ direction in the plaquette. Starting from  Eqs.(\ref{partition}-\ref{T(k)})
we obtain the following expression for  the Green's function of the lattice 
\be\label{gsymclu}
G(\omega,{\bf k})=\frac{1}{4}\sum_{ a,b} \tilde g_{ab}(\omega,{\bf  k})e^{i{\bf k}\cdot ({\bf r}_a-{\bf r}_b)}\ee
where $a$ and $b$ enumerate the sites in the plaquette
and  $\tilde g_{ab}(\omega,{\bf  k})$ is a  $2\times2$ matrix in spin-space denoting the single-particle Green's function of $H$ to lowest order in the interplaquette hopping $V$ \cite{pairault1998} {\sl i.e.}
\be\label{gf}
\tilde g(\omega,{\bf k})=\lf[g^{-1}_{\rm loc}(\omega)-\tilde T[{\bf k}]\rg]^{-1}
\ee
where  $g_{\rm loc}(\omega)$ is the exact plaquette's Green's function.
The overall structure of the spectrum can be then deduced from  the spectral function, $A_0(\omega, {\bf k})$, defined as  $A_0(\omega, {\bf k})=-\frac{1}{2\pi}{\rm Im}\,{\rm Tr}\lf[G(\omega,{\bf k} )\rg]$  while the spin-polarization of the states can be described using the spin-projected spectral function $A_\mu(\omega, {\bf k})=-\frac{1}{2\pi}{\rm Im}\,{\rm Tr}\lf[G(\omega,{\bf k} )\sigma_\mu\rg]$ with $\mu=x,y,z$.

 \begin{figure*}
\begin{center}
\includegraphics[width=\textwidth]{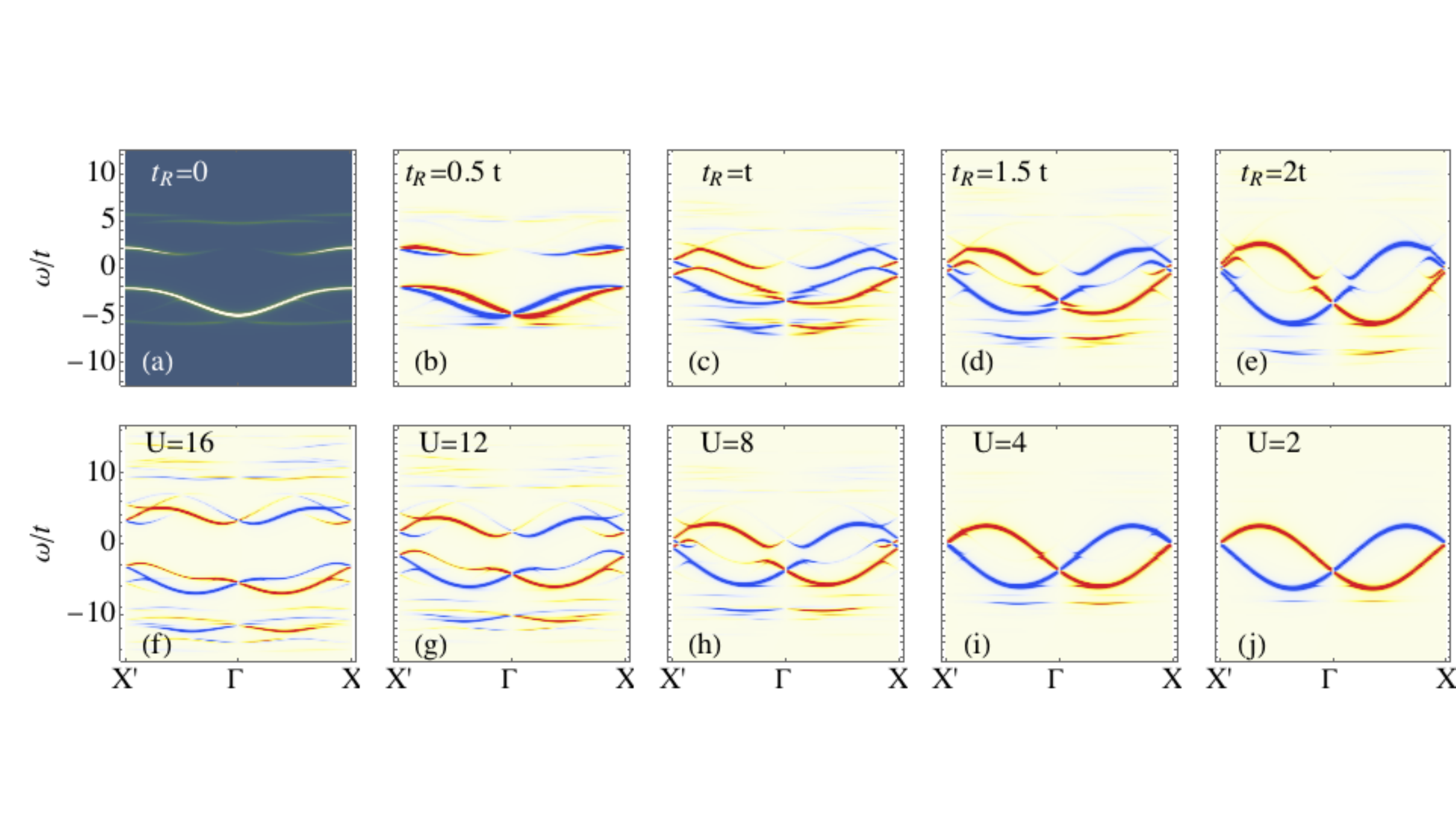}
\caption{Evolution of the spectrum across the Mott transition along the axis $k_x$, red and blue colors indicate positive and negative $y$-polarization.  Mott insulating phases for  weak and strong spin-orbit are shown on the upper and lower leftmost panels while metallic phases for, respectively, strong and weak interaction  are shown on the rightmost panels.  Specifically, upper  panels, (a-e), show  $A_y(\omega, {\bf k})$ for $t_R\in[0,\,2t]$ and $U=6t$ while lower panels, (f-j), show $A_y(\omega, {\bf k})$ for $t_R=2t$ and $U\in[16t,\,2t]$.}\label{Fig-Mott-transition}
\end{center}
\end{figure*} 
\section{Metal-insulator transition in the presence of Rashba SOC}
\label{sec-results}


We start by  presenting  the evolution of the spectrum along two representative lines in the space of parameters which cross the insulator-to-metal transition starting from the insulating solution.  In particular, the first row of Fig.\ref{Fig-Mott-transition} (Panels (a-e)) shows results for fixed $U=6 t$ and increasing values of the Rashba coupling $t_R$, while in the second row (Panels (f-j)) we fix a moderately large value of the RSOC $t_R=2 t$ and we vary $U$ ranging from $U=16 t$ to $U=2t$. To trace the modifications and merging of the different bands we focus on the structure of the spectrum along the line $X'$-$\Gamma$-$X$, with $X=(\pi,0)$, $\Gamma=(0,0)$ and $X'=(-\pi,0)$ and we plot the $\sigma_y$-component of the spectral function, $A_{y}(\omega,{\bf k})$. 

In the absence of RSOC ($t_R=0$), (Fig. \ref{Fig-Mott-transition}a), CPT yields  a spectral functions featuring two well-defined  Hubbard bands along with two satellite bands, similar results were obtained in Ref.[\onlinecite{senechal2000}]
As we switch on a small Rashba coupling, the Hubbard bands acquire an helical structure, as shown in Fig. \ref{Fig-Mott-transition}b and discussed in more details in Appendix A.
A further increase of $t_R$ then induces a transition to a metallic state, (Fig.\ref{Fig-Mott-transition}(c-d)) and, in the limit of large $t_R$ the spectrum strongly resembles the non-interacting one Figs.\ref{Fig-Mott-transition}(e,i,j). Therefore, as $t_R$ goes from $0.5 t$ to $2 t$ the system undergoes a transition from a Mott insulator with spin-split Hubbard bands  to a Rashba metal despite the interaction is unchanged to $U=6t$.

On the other hand, if we start from a Rashba metal with a large value of Rashba SOC, shown in Fig.\ref{Fig-Mott-transition}(h), and we increase Hubbard interaction strength we can drive the system towards an insulating phase, 
 realized at large $t_R$ and strong interaction, that  differs significantly from  the weak-SOC  Mott-insulator shown in Fig. \ref{Fig-Mott-transition}(b). As one can see in Fig. \ref{Fig-Mott-transition}(f), where we show the spectral function for $U=16 t$ and $t_R=2t$,  this strong-SOC insulating phase  is  characterized by a flat spectrum with a very weak ${\bf k}$ dependence.
When $U$ is reduced, the non-interacting band-structure is recovered by merging the outer branches of the two bands, {\sl i.e.} the positive helicity branch of the upper Hubbard band and the negative helicity branch of the lower Hubbard band. The inner branches progressively disappear across the transition as shown in   Fig. \ref{Fig-Mott-transition}(f-j).

\subsection{Mott-Rashba insulators}\label{sec-ins}
 \begin{figure}[t!]
\begin{center}
\includegraphics[width=0.45\textwidth]{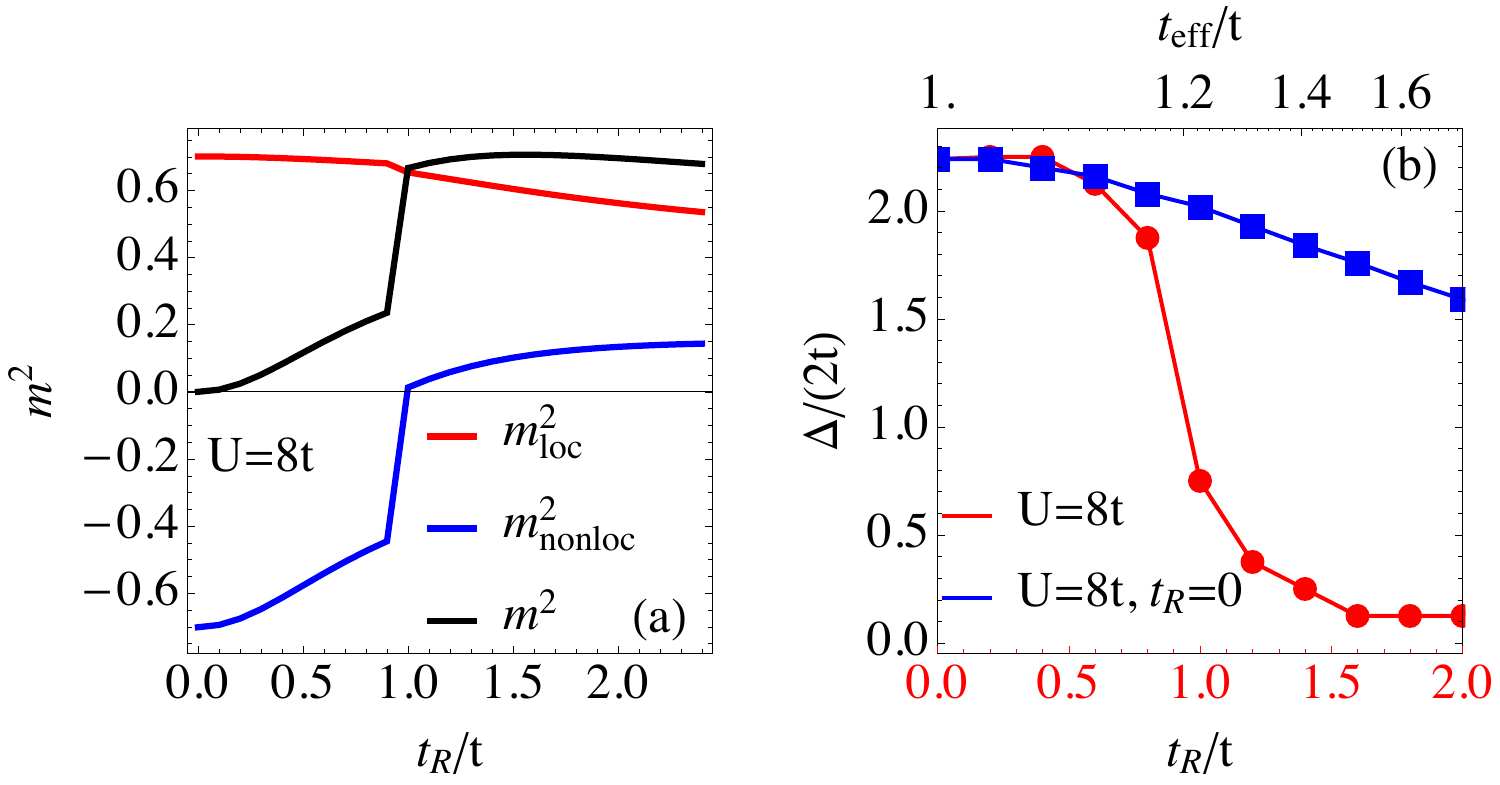}
\caption{(a) Local and non-local magnetization as defined by Eqs.(\ref{sloc}-\ref{snonloc}) as a function of $t_R$ for $U=8 t$. (b)Red line: Mott gap as a function of the ratio $t_R/t$ shown on the lower horizontal scale for $U=8t$.  Blue line: Mott gap for $t_R=0$, $U=8t$ and a diagonal tunneling amplitude $t_{\rm eff}$, shown on the upper horizontal scale. }\label{fig-ssquare}
\end{center}
\end{figure}

To elucidate
the nature of the insulating phases realized at small and large $t_R$ we recall that while at $t_R=0$ Lieb's theorem\cite{lieb1989} prescribes  zero total magnetic moment, {\sl i.e. } 
\be \la S^2\ra=\la S^2_{x}+S^2_{y}+S^2_{z}\ra=0\ee
 with $S_\mu=\sum_ic^\dag_i\sigma^\mu c_i$,
 at $t_R\neq 0$ the breaking of the SU(2)
spin symmetry removes the constraint on $S^2$ that acquires a finite value.
 It is then useful to cast  the average squared magnetization, $m^2=\la S^2\ra/L$, where L is the number of sites, as the sum of a local and a non-local contribution, 
\be m^2=m^2_{\rm loc}+m^2_{\rm nonloc}.\ee
 In the above equation $m^2_{\rm loc}$ quantifies the local magnetic moment and it can be easily related to the average double occupancy,  $d=1/L \sum_i \la n_{i\uparrow}n_{i\downarrow}\rangle$ and to the density per site $n$,
\be m^2_{\rm loc} =\frac{1}{L}\sum_{\mu,i}\la(c^\dag_i\sigma^\mu c_i)^2\ra=3(n-2d)\label{sloc}\ee
while $m^2_{\rm nonloc}$ characterizes the spin-spin correlation between different sites:
\be m^2_{\rm nonloc} =\frac{1}{L}\sum_{\mu,i\neq j}\la(c^\dag_i\sigma^\mu c_i)(c^\dag_j\sigma^\mu c_j)\ra.\label{snonloc}\ee
 From Eqs. (\ref{sloc}) and (\ref{snonloc}) it follows that in the standard Hubbard model with $t_R=0$ the decrease of the double occupancy driven by the interaction is unavoidably associated to the creation of negative non-local spin correlations; the constraint  $m^2=0$ indeed implies $m^2_{\rm nonloc} =-m^2_{\rm loc}$. 
On the other hand, the presence of a finite Rashba coupling removes the constraint $m^2=0$  and it allows $m^2_{\rm nonloc}$ and $m^2_{\rm loc}$ to vary independently.
 
This is clearly shown in Fig.\ref{fig-ssquare}(a) where we plot the local and non-local magnetization as well as their sum, approximated using the single-cluster ground-state, as a function of $t_R/t$ for $U=8t$.
There are two fundamentally different regimes: the first, realized at small $t_R$, is characterized by  $m^2_{\rm nonloc}<0$ corresponding to predominant antiferromagnetic (AF) correlations, in this regime we recover the standard result 
$\la S^2\ra=0$ at $t_R=0$, the second regime, realized at large $t_R$, is instead characterized by $m^2_{\rm nonloc}\gtrsim0$ indicating predominantly helical (HL) correlations.\cite{FN1}

A natural question that arises here is whether the onset of the HL  regime affects the metal-insulator transition. To answer this question,  in Fig.\ref{fig-ssquare}(b) we plot the Mott gap as a function of $t_R/t$ for $U=8t$ and we compare it with the gap of a Hubbard model having the same bandwidth but no RSOC, {\sl i.e.} with a standard square lattice Hubbard model having an effective  spin-diagonal hopping amplitude,  $t_{\rm eff}=\sqrt{t+t_R^2/2}$.\cite{FN2}
 We see that the onset of the HL regime brings about a change in the behavior of the 
 Mott gap that becomes strongly sensitive to the ratio of $t_R/t$ and it falls  rapidly to zero.  On the contrary, in the absence of Rashba SOC, we obtain a much weaker dependence on the ratio $U/t_{\rm eff}$.
 Including dynamic correlations neglected by CPT, or increasing the cluster size will probably reduce the gap but it will not qualitatively modify its behavior, as briefly outlined in Appendix \ref{LSF}.
 \noindent These results suggest that the metal-insulator transition is then not simply driven by the decrease of the  double occupancy, controlled by the ratio $U/t_{\rm eff}$, but it also depends on the strength and nature  of  non-local spin correlations.
 Further evidences in this direction may be found in appendix \ref{sec-tNNN} where we compare the effects Rashba coupling  and of next-nearest-neighbour tunneling on the charge gap.
There we show that, although the latter kind of hopping destroys the nesting and reduces the density of states at the Fermi level, its overall effect on the charge gap is rather different from that of RSOC for moderate and large values.
 The presence of Rashba spin-orbit coupling thus seems to  favor metallic phases  and to introduce new  screening mechanisms. 
%
%
A somewhat similar situation is realized in the presence of Hund's coupling in multiorbital systems where more than one orbital is available on every lattice site. Here the role of non-local spin correlations is played by the orbital magnetization.\cite{isidori2019}
Our analysis also shows that  the effect of Rashba coupling differs from that of  an external constant magnetic field despite a superficial similarity. 
In this simpler case, the total magnetic moment increases as a function of the magnetic field strength, mostly due to an increase of the local magnetization,  while Rashba SOC instead mostly affects the non-local magnetization. This leads to a basic and important difference in the limit of very large couplings. Obviously very large magnetic fields  drive the system towards a band insulator\cite{laloux} while a very large Rashba SOC leads to a semimetallic state,
as we discuss in more details in the following Section.

Before proceeding further let us add a technical remark.
We extract the gap from the density of states. In order to obtain a  numerical estimate, we use twice the energy of the first point where the density of states changes curvature from positive to negative (the factor two comes from the symmetry around zero energy of the DOS). As one can easily understand by looking at Fig.\ref{fig-DOS} for each value of $t_R$, this procedure yields the ``true'' gap and not the pseudogap.

%
%
%
%
 %

\subsection{Correlated Rashba metal}  \label{sec-met}

 \begin{figure}[t!]
\begin{center}
\includegraphics[width=0.5\textwidth]{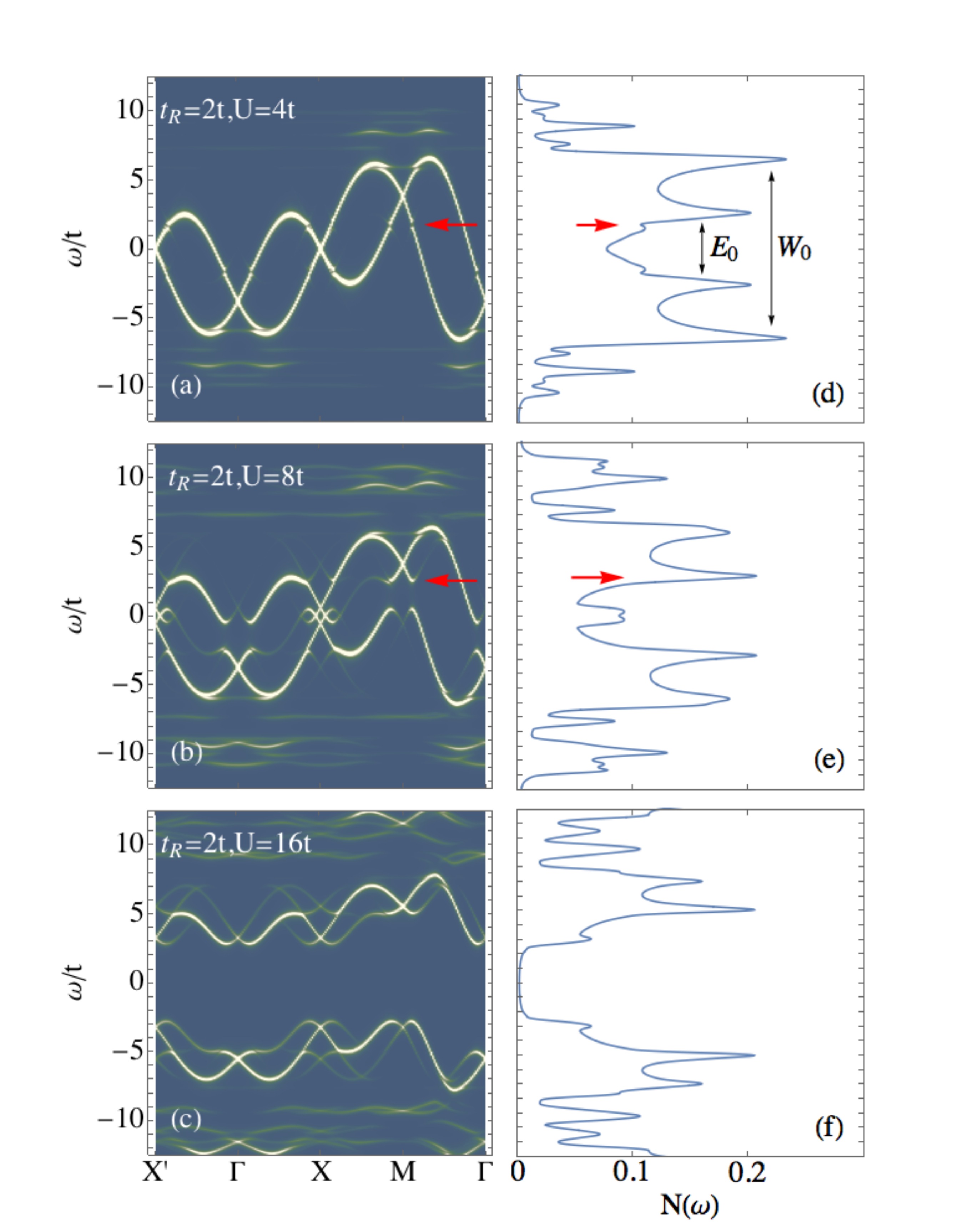}
\caption{Evolution of the spectral function $A_0(\omega,{\bf k})$ across the Mott transition for $t_R=2t$ and $U\in[4t,\,16t]$, Lorentzian broadening $\eta=0.1t$ in the spectral function plots and $\eta=0.25 t$ in the DOS plots. The red arrows in panels (a,d) signal rifts in the spectral function and the corresponding peaks in the DOS.}\label{fig-DOS} 
\end{center}
\end{figure} 

 As one can see in Fig.\ref{Fig-Mott-transition}(d,h), a peculiarity of the Mott transition in the presence of Rashba SOC is that at the transition the spectrum remains ungapped around the X and $Y=(0,\pi)$ points while a pseudogap appears around  $\Gamma$ and $M=(\pi,\pi)$ points in the first Brillouin zone.  This behavior can be qualitatively understood considering that the Rashba spin-orbit coupling introduces  Fermi-level Dirac crossings at the X and Y points.  
Low-energy fermions  with momentum close to the  X and Y points therefore  behave as nodal fermions and they are more robust  to interaction effects as compared to the standard fermions present around $\Gamma$ and $M$ points.

\begin{figure}[t!]
\begin{center}
\includegraphics[width=0.23\textwidth]{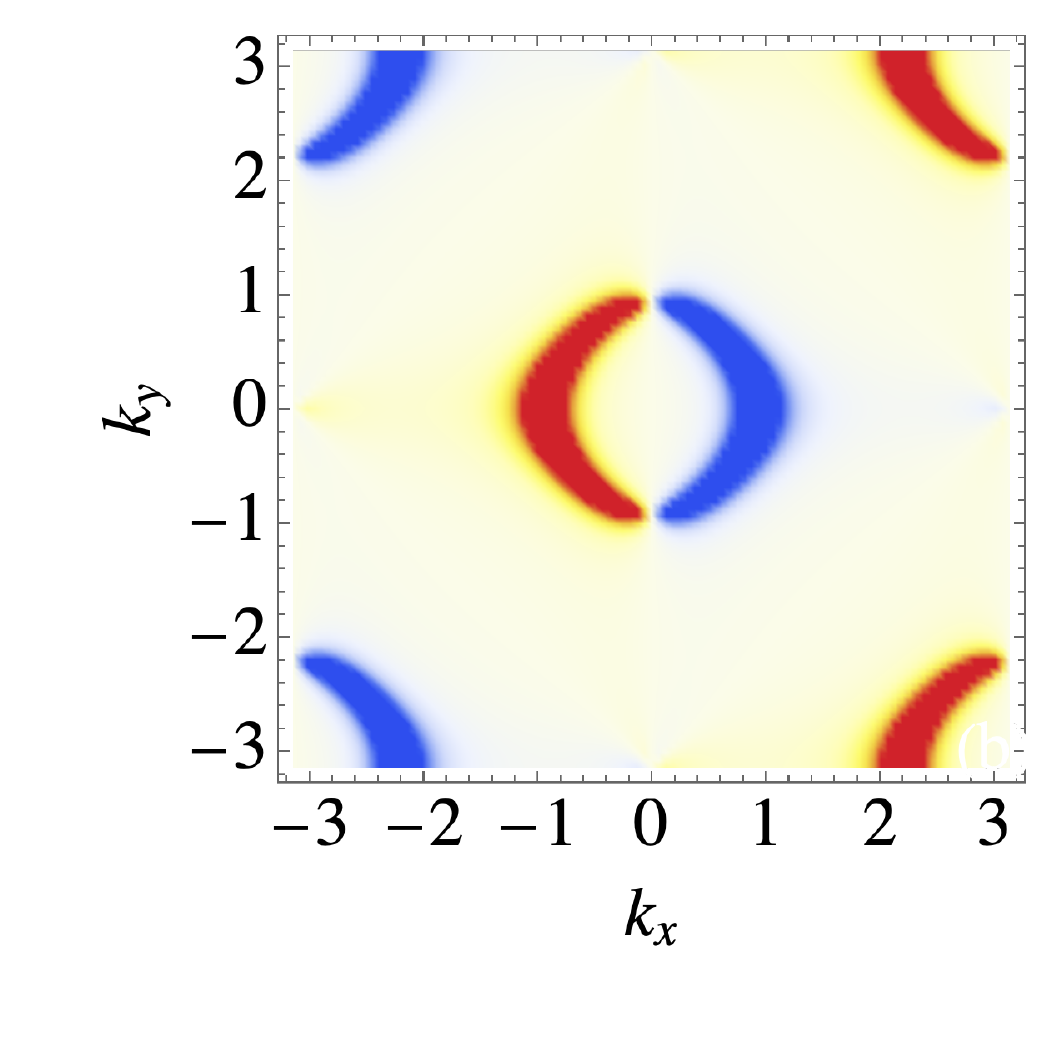}
\includegraphics[width=0.23\textwidth]{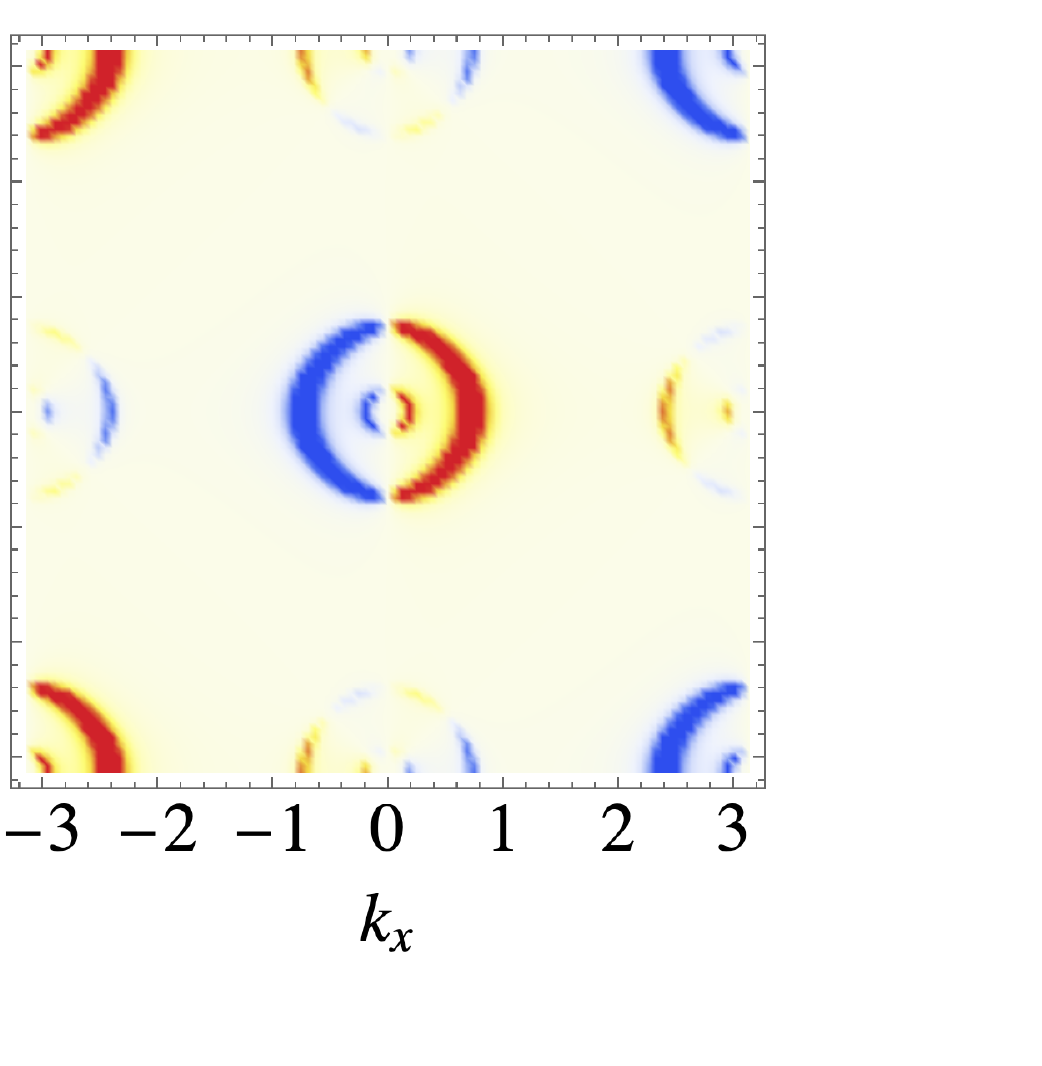}
\caption{ Spin-polarized spectral function,  $A_y(\omega, {\bf k})$, at the Fermi level, $\omega=0$,  for $t_R=2t\, {\rm and}\,U=0$ (left)  and  $t_R=2t\, {\rm and}\,U=8t$ (right).}\label{fig-FS}
\end{center}
\end{figure}

Signatures of the presence of two kinds of fermionic quasi-particles behavior may be also found in the density of states shown in Fig.\ref{fig-DOS}.
There we see that in the weakly correlated phase, shown in Fig.\ref{fig-DOS}(a), the spectrum bears strong similarities with the non-interacting one. In particular, due to the presence of Rashba coupling, the  Van Hove singularity at $\omega=0$ characteristic of two-dimensional square lattices is  split into two peaks separated by an energy $E_0=4t(\sqrt{1+t_R^2/t^2}-1)$.
At the same time two additional van-Hove singularities appear at the band edges with a distance proportional to the non-interacting bandwidth $W_0=8t\sqrt{1+t_R^2/(2t^2)}$.
The signatures of correlation, in this case,  are the transfer of spectral weight at large $\omega$ associated with spin-wave excitations and  the appearance of rifts at the points $\Gamma$ and $M$ yielding  the two small peaks close to the Fermi level indicated with a red arrow in Fig. \ref{fig-DOS}(a).
As we increase interaction, the rifts evolve into a pseudogap  (Fig. \ref{fig-DOS}(b)) and two additional  peaks appear associated with the opening of the gap at the  points $X$ and $Y$. This  intermediate regime  is characterized by a Fermi surface shown in Fig. \ref{fig-FS} consisting of two circles centered at the $\Gamma$ and $M$ points with  two additional pockets appearing at the points $X$ and $Y$ and displaying a complex spin texture reminiscent of that recently found in Ref.[\onlinecite{gotlieb2018}].

Fig. \ref{fig-DOS}(c) shows the DOS in the insulating phase. In this phase the four-peak structure characteristic of the non-interacting Rashba metal on the square lattice is reproduced in each of the two Hubbard bands.
However, the changes in the bands dispersion induced by Hubbard interaction, clearly visible in Fig. \ref{fig-DOS}(c), demonstrate a non-trivial renormalization of the different contributions to the kinetic energy.    We notice that  cluster-Dynamical Mean-Field Theory studies of the two-dimensional Hubbard models have shown similar physics where the interactions give rise to different renormalizations of nearest-neighbor and further-range hoppings, leading to renormalizations of the Fermi surface \cite{Civelli1} and appearance of the pseudogap and the superconducting gap \cite{Civelli2}.

\section{conclusions}
In this manuscript we have studied the interplay between a sizable Rashba spin-orbit coupling and Hubbard-like local interactions in a two-dimensional lattice model. 
We find that the presence of RSOC deeply affects the Mott transition and it has two crucial effects: (i) it makes local and non-local magnetization independent, in contrast with the pure Hubbard model, where the formation of large local magnetic moments leads to negative non-local spin correlations and (ii) it introduces robust nodal quasiparticles.  As a consequence, the RSOC strongly favors metallic phases, turning a Mott insulator into a Rashba metal through a transition which can not be described in terms of an effective standard Hubbard model with a renormalized kinetic term.  The spectral properties reveal different mechanism in which the insulator transforms into a metal and underline a strong sensitivity of the spectral and transport properties on the value of the ratio between the RSOC and the standard hopping amplitude. Our results provide the community with simple and practical information about the strong effect of the spin-orbit coupling on Mott localization, which can be used as the cornerstone for the study of systems featuring simultaneously large RSOC and large Hubbard $U$ and as guidelines to tune and tailor the properties of these systems. To be concrete, let us discuss in more detail the connection with two recent experimental works which are pioneering the field of strongly correlated Rashba systems, reported in Refs. [\onlinecite{tresca2018}] and [\onlinecite{gotlieb2018}].

In Ref.[\onlinecite{tresca2018}]  Tresca {\sl et al.} investigate the 1/3 monolayer $\alpha$-Pb/Si(111) by scanning tunneling microscopy
finding a metallic ground state in sharp contrast  to what happens {\sl e.g.}  in Sn on Si(111) that is a Mott insulator.\cite{profeta2007}
By detailed first principle calculations including on equal footing relativistic and correlation effects they show that a peculiar feature of the former system is the strong spin-orbit coupling.
Despite our simple single-band square-lattice Rashba-Hubbard model cannot fully describe the complexity of  $\alpha$-Pb/Si(111), a simple analysis of the results of Ref.[\onlinecite{tresca2018}] 
shows that their parameters are in a regime where Pauli screening is relevant. Surface-band electrons  indeed experience a very strong Hubbard interaction and a significant spin-orbit splitting of about 25\% of the bandwidth which in our model would correspond to $t_R\sim1.5 t$.

In Ref.  [\onlinecite{gotlieb2018}], the author unveil the spin-texture of the Fermi surface of the cuprate superconductor Bi2212 by spin-resolved ARPES. We already noticed that a similar spin-texture arises within our model. 

The present work also triggers a number of interesting questions concerning the effects of inversion symmetry breaking and Rashba spin-orbit coupling on  superconductivity in strongly correlated systems. In fact, while it is well-known that Rashba coupling may enhance electron-phonon superconductivity \cite{cappelluti2007}  the effects of spin-orbit coupling on  high-$T_c$ and unconventional superconductors are still to a large extent unknown \cite{fischer2018}.  

\section{acknowledgements}
We are grateful to A. Amaricci  for precious discussions. We acknowledge support from the H2020 Framework Programme, under ERC Advanced GA No. 692670 FIRSTORM and from MIUR PRIN 2015 (Prot. 2015C5SEJJ001) and SISSA/CNR project `Superconductivity, Ferroelectricity and Magnetism in bad metals" (Prot. 232/2015).

\appendix

\section{Cluster perturbation theory in the presence of Rashba spin-orbit coupling} \label{app-A}
The general idea of cluster perturbation theory\cite{senechal2002,pairault2000} (CPT) is to construct an approximate solution for interacting lattice models starting from the exact solution of individual clusters of sites by means of a perturbation theory in the intercluster hopping.\\

This statement already contains essential information concerning CPT, {\sl  i.e.} 
(i) it is a perturbation theory in the hopping and as such it is delicate since the reference  system is interacting;
(ii) it is controlled to some extent by the cluster size, $L_c$, since it becomes trivially exact in the limit $L_c\rightarrow \infty$ and, as shown {\sl e.g.} in Ref.[\onlinecite{sarker1988}] it reduces to the Hubbard I approximation for $L_c=1$.\\

The derivation of CPT equations can be done in several ways by means of diagrammatic perturbation theory\cite{pothoff2014} or path integral approaches.\cite{pairault2000}
Here we  briefly outline the path integral derivation illustrating  the differences arising due to the presence of Rashba spin-orbit coupling\cite{rashba1959,WinklerSpinOrbitCoupling} (SOC).

\subsection{Model and Hamiltonian partitioning}

\label{app-model}
The model is described by the following Hamiltonian:
\be
H=H_0+H_U
\ee 
where $H_U$ denotes the Hubbard interaction, 
\be
H_U=U\sum_{i}n_{i\uparrow}n_{i\downarrow}
\ee
while $H_0$ indicates the non-interacting Hamiltonian and it includes a spin-diagonal hopping proportional to $t$ and a spin-dependent hopping quantified by $t_R$ associated with Rashba spin-orbit coupling. Introducing the spinor creation and annihilation operators $c^\dag_i=(c^\dag_{i\uparrow},c^\dag_{i\downarrow})$ and $c_i=(c_{i\uparrow},c_{i\downarrow})$, the Hamiltonian $H_0$ can be cast al follows
\be
H_0=\sum_{ij}c^\dag_i\lf[t\, \sigma_0\Delta_{ij}+t_R (\vec \alpha_{ij}\times\vec \sigma)_z\rg]c_j
\ee
where $i$ enumerates the sites of 2D square lattice, $Z_\gamma$, $\vec\sigma$ denotes the vector of Pauli matrices,  $\vec \alpha=\lf(\alpha_{ij}^x,\alpha_{ij}^y,0\rg)$ and $\alpha_{ij}^\mu=i(\delta_{ij+\eta_\mu}-\delta_{ij-\eta_\mu})$ and $\Delta_{ij}=\sum_{\mu}(\delta_{ij+\eta_\mu}+\delta_{ij-\eta_\mu})$ with $\eta_x$ and $\eta_y$ denoting the unit vectors (1,0) and (0,1).

The first step to construct a CPT consists in  tiling the lattice with small clusters whose Hamiltonian can be diagonalized exactly. As clusters we use $2\times2$ plaquettes: this choice has preserves the symmetries of the lattice and allows to easily account for  the chiral structure of Rashba SOC.   Furthermore, thanks to the small clusters size, it allows the investigation of a wide region of parameters with small computational efforts.

The clusters superlattice,  $Z_\Gamma$,  is defined  by decomposing the vectors of the original lattice, $Z_\gamma$,  as follows
 \be {\bf r}_i={\bf r}_m+{\bf r}_a\label{map}\ee 
 where ${\bf r}_m$ denotes the position of the lowermost left site of each plaquette while ${\bf r}_a$ indicates the position of the site inside the plaquette. Therefore, ${\bf r}_m=(m_x\,a_x,m_y\,a_y)$ with $m_x$ and $m_y$ even numbers and   ${\bf r}_a=(\beta_x\,a_x,\beta_y\,a_y)$ with $\beta_x,\beta_y\in[0,1]$ and $a_\mu$ denote the primitive lattice  vectors.

\begin{figure}[t!]
\begin{center}
\includegraphics[width=0.35\textwidth]{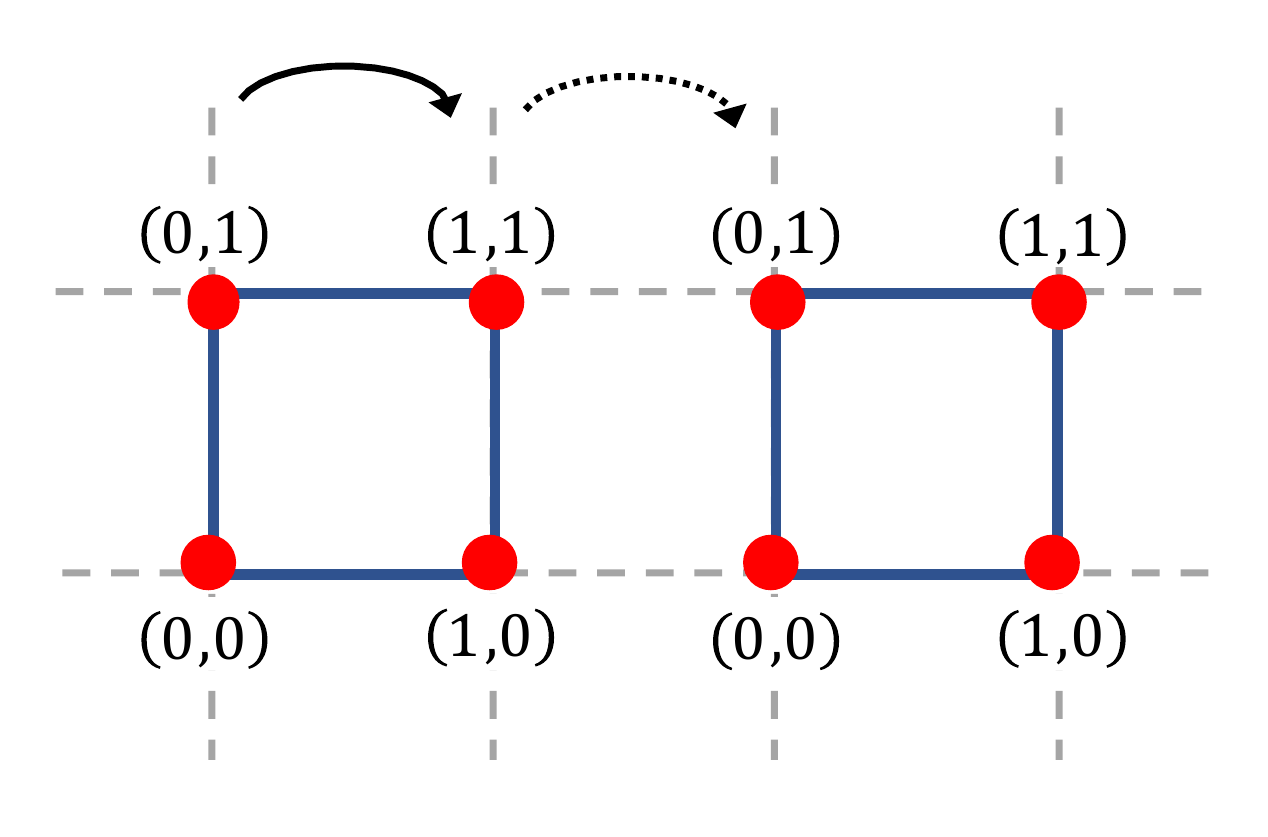}
\caption{The solid and dotted lines show examples of forward intra- and inter- plaquette tunneling along $x$.}\label{fig:interplaquettetunneling}
\end{center}
\end{figure}

Using the map defined in Eq.\eqref{map} we can easily rewrite the hopping Hamiltonian as  follows:
\be
H_0=-\sum_{m,n,a,b}c^\dag_{am}\, T_{am,bn}\,c_{bn}.
\ee
where $T_{am,bn}=T^x_{am,bn}+T^y_{am,bn}$  is a matrix in spin-space that describes the hopping of one electron from site $b$ of cluster $n$ to site $a$ of cluster $m$.
The structure of the hopping matrix, $T_{am,bn}$,  can be most easily understood 
by looking at Fig.\ref{fig:interplaquettetunneling}. There we see that,  given the map defined in Eq.\eqref{map}, forward (backward) interplaquette tunneling is associated with backward (forward) translation of the intracluster indices. In the presence of Rashba SOC this leads to the following expression for the tunneling along $x$
\bea 
T^{x}_{am,bn}\!\!\!\!&=&\!\!\!\lf(\delta_{m,n+2\eta_x}\delta_{a,b-\eta_x}+\delta_{m,n}\delta_{a,b+\eta_x}\rg)\![t\sigma_0+i\,t_R\sigma_y]\!+\nn\\ & &\!\!\!\!\!\!+\lf(\delta_{m,n-2\eta_x}\delta_{a,b+\eta_x}+\delta_{m,n}\delta_{a,b-\eta_x}\rg)[t\sigma_0-i\,t_R\sigma_y]\nn\\
\eea
and for the tunneling along $y$
\bea 
T^{y}_{am,bn}\!\!\!\!&=&\!\!\!\lf(\delta_{m,n+2\eta_y}\delta_{a,b-\eta_y}+\delta_{m,n}\delta_{a,b+\eta_y}\rg)\![t\sigma_0-i\,t_R\sigma_x]\!+\nn\\ & &\!\!\!\!\!\!+\lf(\delta_{m,n-2\eta_y}\delta_{a,b+\eta_y}+\delta_{m,n}\delta_{a,b-\eta_y}\rg)[t\sigma_0+i\,t_R\sigma_x].\nn\\
\eea

Starting from the above equations it is straightforward to partition the Hamiltonian as 
\be
H=H_{\rm loc}+\hat V
\ee
where $V$ describes the inter-cluster hopping while $H_{\rm loc}$ contains all intra-cluster terms.

Specifically, arranging the creation and annihilation operators on a given cluster in a vectorial form, $\hat V$ can be written as 
\be \hat V=-\sum_{m\neq n}C^\dag_{m}\, V_{m,n}\,C_{n}\ee
%
with $V_{m,n}=V^x_{m,n}+V^y_{m,n}$ and {\sl e.g.} $V^x_{m,n}$ given by
\bea V^x_{m,n}\!\!&=&\!\!\delta_{m,n-2\eta_x}\tau^+_x\!\otimes(t\sigma_0-i\,t_R\sigma_y)+\nn\\
& & +\delta_{m,n+2\eta_x}\tau^-_x\!\otimes(t\sigma_0+i\,t_R\sigma_y)\eea
where $\tau^+_\mu$ ($\tau^-_\mu$ ) denote a single-site forward (backward) translation in the $\mu$ direction of the intra-cluster indices. Similarly, $H_{\rm loc}$ can be cast as 
\be H_{\rm loc}=-\sum_{m}C^\dag_{m} T'\,C_{m}+H_U\ee
with %
\bea \!\!\!T'\!\!&=&\!\!\tau^+_x\!\otimes(t\sigma_0+i\,t_R\sigma_y)+\tau^-_x\!\otimes(t\sigma_0-i\,t_R\sigma_y)+\nn\\
&&\!\!\!\!\!\!\!+\tau^+_y\!\otimes(t\sigma_0-i\,t_R\sigma_x)+\tau^-_y\!\otimes(t\sigma_0+\,t_R\sigma_x).\eea
From the above equations we see that the presence of Rashba coupling leads to a  redefinition of the local Hamiltonian and of the inter-cluster hopping. Both these terms acquire a spinorial structure and are described by matrices of dimension $2L_c$ where $L_c$ denotes the cluster size.
\subsection{Path integral derivation of CPT equations}
The purpose of this section is to present a schematic derivation of the CPT relation between cluster and lattice  Green's functions, Eq. (\ref{gsymclu}-\ref{gf}) of section \ref{sec-model}, using a path integral approach. \\
We start by introducing  the vectors of Grassman fields $\Gamma_{m}$ and $\Gamma^*_{m}$ corresponding to the vectors of  fermion operators $C_{m}$ and $C^\dag_{m}$ and 
we write the partition  function as 
\bea
Z&=&\!\!\int {\cal D}(\Gamma^*_m,\Gamma_{m})\exp\Big[{-\int_0^\beta\big(\Gamma^*_{m}\partial_\tau\Gamma_{m}+H_{\rm loc}\big){\rm d}\tau}\Big]\cdot\nn\\
& &\cdot \exp\Big[{\int_0^\beta \Gamma^*_{m}\, V_{mn}\,\Gamma_{n}{\rm d}\tau}\Big]
\eea
where we follow the notation of Ref.[\onlinecite{negele}] and the sum over repeated indices in the exponent is implied.
Following the route outlined in Refs.[\onlinecite{pairault2000,senechal2002,sarker1988}], we perform a Grassmannian Hubbard-Stratonovich transformation and  recast the partition function as 
a Gaussian integral over the auxiliary Grassman fields, $\Psi_m$ and $\Psi^*_m$. By doing so we obtain
\begin{widetext}
\be
Z={{\rm det} V}\int {\cal D}(\Gamma^*_m,\Gamma_{m}){\cal D}(\Psi^*_m ,\Psi_{m}) \exp\Big[{-\int_0^\beta\!\big(\Gamma^*_{n}\partial_\tau\Gamma_{n}+H_{\rm loc}\big){\rm d}\tau}\Big]
\exp\Big[{\int_0^\beta \Big(\Psi^*_{m}\, V^{-1}_{mn}\,\Psi_{n}+\Psi^*_n\Gamma_n+\Gamma^*_n\Psi_n\big){\rm d}\tau}\Big].
\ee
\end{widetext}
We see that the $\Gamma$-fields action is completely local,  this allows to factorize the $\Gamma$-integral 
and arrive at the following expression for the partition function
\bea
Z&=&\!\!{{\rm det} V}\,Z_0\!\int\! {\cal D}(\Psi^*_m,\Psi_{m} )
\exp\Big[{\int_0^\beta \Psi^*_{m}\, V^{-1}_{mn}\,\Psi_{n}{\rm d}\tau}\Big]\cdot \nn\\
& & \cdot\exp\Big[\sum_m S_m( \Psi^*_{m},\Psi_{m})\Big]
\eea
where $Z_0=\Pi_m z^0_{m}$ denotes the local partition function, {\sl i.e.} $z^{0}_m=\int {\cal D}(\Gamma \Gamma^*)e^{-S_{\rm loc}}$ with $S_{\rm loc}=\int_0^\beta\!\big(\Gamma^*_{n}\partial_\tau\Gamma_{n}+H_{\rm loc}\big){\rm d}\tau$,  while $S_m( \Psi^*_{m},\Psi_{m})$ is defined as follows
\be\label{eq:sm0}
S_m( \Psi^*_{m},\Psi_{m})=\log\big(\big\la e^{-\!\int_0^\beta (\Psi^*_m\Gamma_m+\Gamma_m^*\Psi_m){\rm d}\tau}\big\ra_{\! \rm 0}\big)
\ee
where the subscript $0$ indicates averages with a local  statistical weight $\la\ldots\ra_0=1/Z_0\int {\cal D}(\Gamma \Gamma^*)\ldots e^{-S_{\rm loc}(\Gamma, \Gamma^*)}$.
Starting from Eq.\eqref{eq:sm0} and expanding the exponential yields a complicate interacting theory for the auxiliary fields: CPT consists in keeping only the first order of this expansion setting
\be\label{eq:sm}
S_m( \Psi^*_{m},\Psi_{m})\!\simeq\!-\!\!\int_0^\beta\!\! \!d\tau_1d\tau_2\,\psi^*_m(\tau_1)\,G_{\rm loc}(\tau_1-\tau_2)\,\psi_m(\tau_2)
\ee
where $G_{\rm loc}(\tau_1-\tau_2)$ denotes  the cluster Green's function and it is a matrix in the cluster-site and spin indices.
Using Eq. \eqref{eq:sm}, switching to Matsubara frequencies and introducing appropriate source fields, we can easily derive the following relations:
\be
G_{\rm CPT}=V^{-1}+V^{-1}G_{\rm aux}V^{-1}
\ee 
where $G_{\rm aux}$  and $G_{\rm CPT}$ denote the $\Psi$'s and $\Gamma$'s Green's functions. 
Eventually, using $G_{\rm aux}=-(V^{-1}+G_{\rm loc})^{-1}$ we get the fundamental equation of CPT for the electronic Green's function:
\be
G_{\rm CPT}=(G_{\rm loc}^{-1}+V)^{-1}.
\ee 
\subsection{Lattice spectral function}
\label{LSF}

Up to this point spin-indices played the same role as cluster-site indices and the effect of Rashba coupling was considered only in the partitioning of the Hamiltonian. The spin structure of the Green's function becomes relevant again when we use CPT to extract physical information on the lattice ground-state.   

Before proceeding further we notice that for single-site cluster the above expression reduces to Hubbard I approximation for the single-particle Green's function.
 Within this approximation the self-energy is independent of spin and momentum and the interacting spectrum consists of four bands with dispersion 
\be
\label{Eigensystem-sc}
\xi_{\bf k\l\nu}=\frac{U+E_{\bf k\l}+\nu\sqrt{U^2+E^2_{\bf k\l}}}{2} 
\ee
%
where  the index $\nu=\pm$ identifies the upper and lower Hubbard band while the index $\lambda=\pm$ indicates the helicity with $E_{\bf k \lambda}$ denoting the dispersion of the helical bands in the absence of interaction, 
\be
\label{Eigensystem}
E_{\bf k\l}= \epsilon_0({\bf k})+2\lambda t_R\sqrt{\sin^2 k_x+\sin^2 k_y}, 
\ee
 where we defined $\epsilon_0({\bf k})=-2 t\lf[\cos(k_x)+\cos(k_y)\rg]$ and we set $a_x=a_y=1$.\\

For finite-size clusters, the different components of the Green's function $G_{\rm CPT}$ are identified by three pairs of indices, $(m,n)$, $(a,b)$ and $(\sigma,\sigma')$, referring respectively to the position of the cluster and the position and the spin of the electron in the cluster. 
In order to arrive to a practical expression for the spectral function of the original lattice model starting from $G_{CPT}$, we switch to momentum space, setting
\be
\hat G_{\rm CPT}({\bf k},{\bf k}')=\frac{1}{L}\!\sum_{a,b,m,n} G_{\rm CPT}\,e^{i\lf(\bf{k}(r_m+r_a)-\bf{k'}(r_n+r_b)\rg)}\ee
where ${\bf k}$ and ${\bf k'}$ belong to the Brillouin zone of $Z_\gamma$.
 Since CPT preserves the superlattice periodicity, the sum over $m,n$ is straightforward and it yields:
\be
G({\bf k},{\bf k}')=\frac{1}{L_c}\!\sum_{a,b}\sum_{\bf K} \delta({\bf k-\bf k}'-{\bf K}) g(\omega,{\bf k})\,e^{i(\bf{k\, r_a-k^\prime\, r_b})}\label{eq:a}\ee
where $K$ belongs to the reciprocal superlattice and $g(\omega,{\bf k})$ is the CPT Green's function in the mixed representation defined by Eq. \eqref{gf} of Section \ref{sec-model}.
\begin{figure}[t!]
\begin{flushleft}
\includegraphics[width=0.5\textwidth]{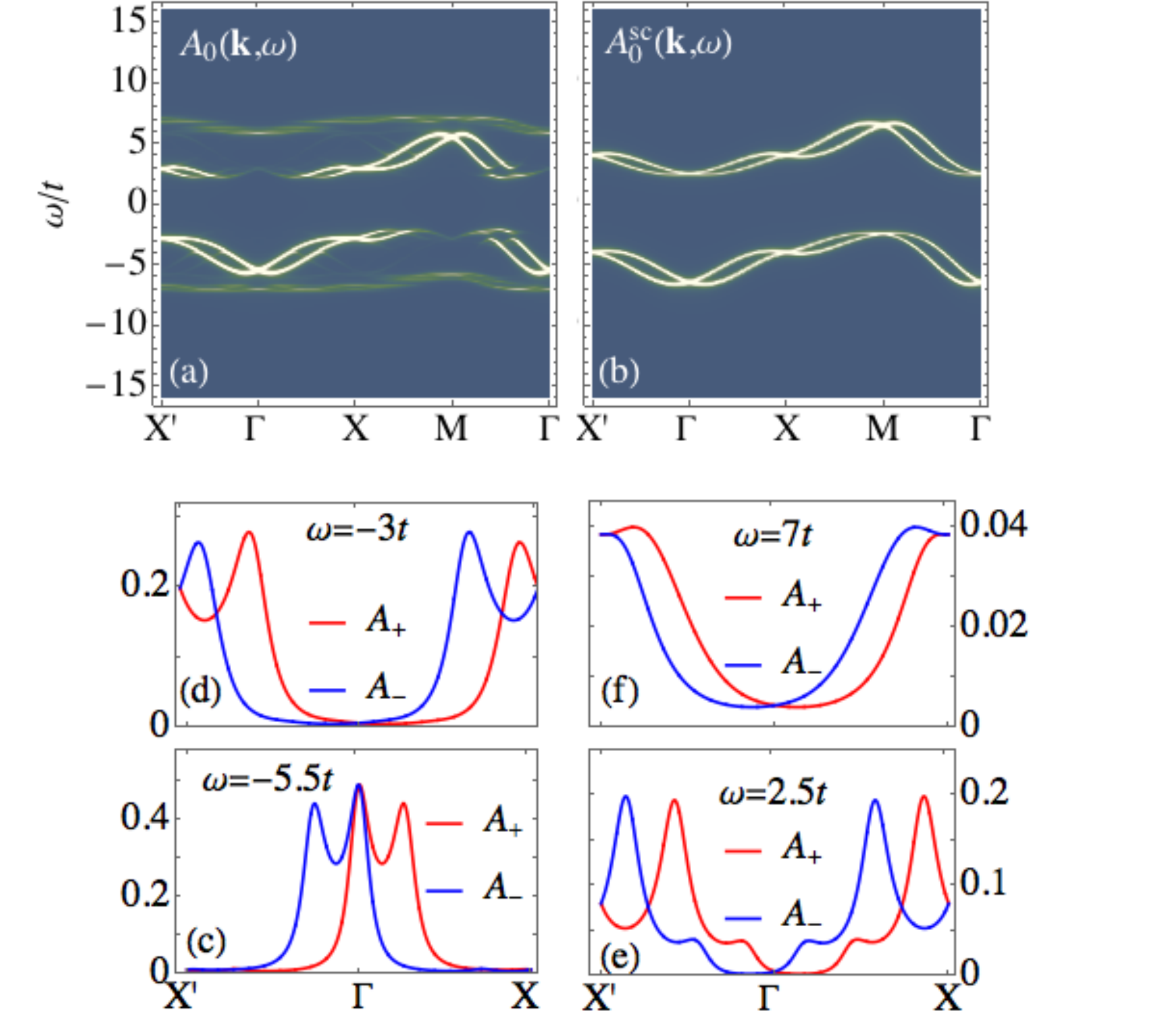}
\caption{(a) Spectral function, $A_0$, of the Hamiltonian $H$ obtained from CPT for $t_R=0.5t$, $U=8t$ and small Lorentzian broadening $\eta=0.1t$ plotted as a function of energy and momentum across the high-symmetry points:  $\Gamma=(0,0)$, $M=(\pi,\pi)$, $X=(\pi,0)$, $X'=(-\pi,0)$.  (b) Spectral function obtained from the strong coupling expansion (Eq.\eqref{Eigensystem-sc}) (c-g) Spin-resolved components of the spectral function along the axis $k_x$, $A_\pm(\omega, {\bf k})=A_0(\omega, {\bf k})\pm A_y(\omega, {\bf k})$, for different values of $\omega$. Parameters values as in (a). }\label{fig-sc-results}
\end{flushleft}
\end{figure} 

Following Ref.[\onlinecite{senechal2002}], we eventually restore the periodicity of the original lattice by keeping only terms with ${\bf k}={\bf k}^\prime$ thus defining the Green's function as 
\be
G_{\sigma\sigma'}(\omega,{\bf k})=\frac{1}{L_c}\sum_{a,b} g_{ab\,\sigma\sigma'}(\omega,{\bf k})e^{i\bf{k}(r_a-r_b)}\label{eq:b}\ee 
where we restored the spin indices.
Comparing Eqs.\eqref{eq:a} and \eqref{eq:b}, one realizes that the periodization prescription of Ref.\onlinecite{senechal2002} can be interpreted as neglecting Umklapp scattering on the cluster superlattice.

A  preliminary understanding of the  effects included in CPT can be gained by looking at Fig. \ref{fig-sc-results}(a-b)  where we compare the CPT spectral function, $A_0(\omega, {\bf k})$
 to that obtained using the Hubbard I approximation, $A^{\rm HI}_0(\omega, {\bf k})$, the latter yields a good qualitative description of the overall structure of the spectrum at small $t_R/t$, but,  featuring a local spin-independent self-energy, it does not capture  the dependence of the spectral weights on ${\bf k}$ and on the spin. 
We notice that in  the large $U$ small $t_R$ limit, a weak spin-orbit coupling induces a finite helical spin-polarization in the paramagnetic Mott insulator. This is apparent in Fig. \ref{fig-sc-results}(c-g), where we plot the spin-resolved spectral function, $A_{\pm}(\omega,{\bf k})$, at different energies for $k_y=0$ and $k_x\in[-\pi,\pi]$. For these ${\bf k}$-values the RSOC behaves as an  effective k-dependent magnetic field pointing along the $y$ axis  and the Green's function can be easily diagonalized yielding $A_{\pm}(\omega,{\bf k})=A_0(\omega,{\bf k})\pm A_y(\omega,{\bf k})$ for the two helicities.

 We remark that, for the standard Hubbard model, a variety of papers have shown that CPT gives a proper account of the most important features of the model. In spite of the small clusters size, the plaquette-perturbation-theory approach used in the present work yields results for the gap that approximately agree with those found using larger 4x4 cluster, see {\sl e.g.} Refs. [\onlinecite{kohno2012,kohno2014}].

\begin{figure}[t!]
\begin{center}
\includegraphics[width=0.5\textwidth]{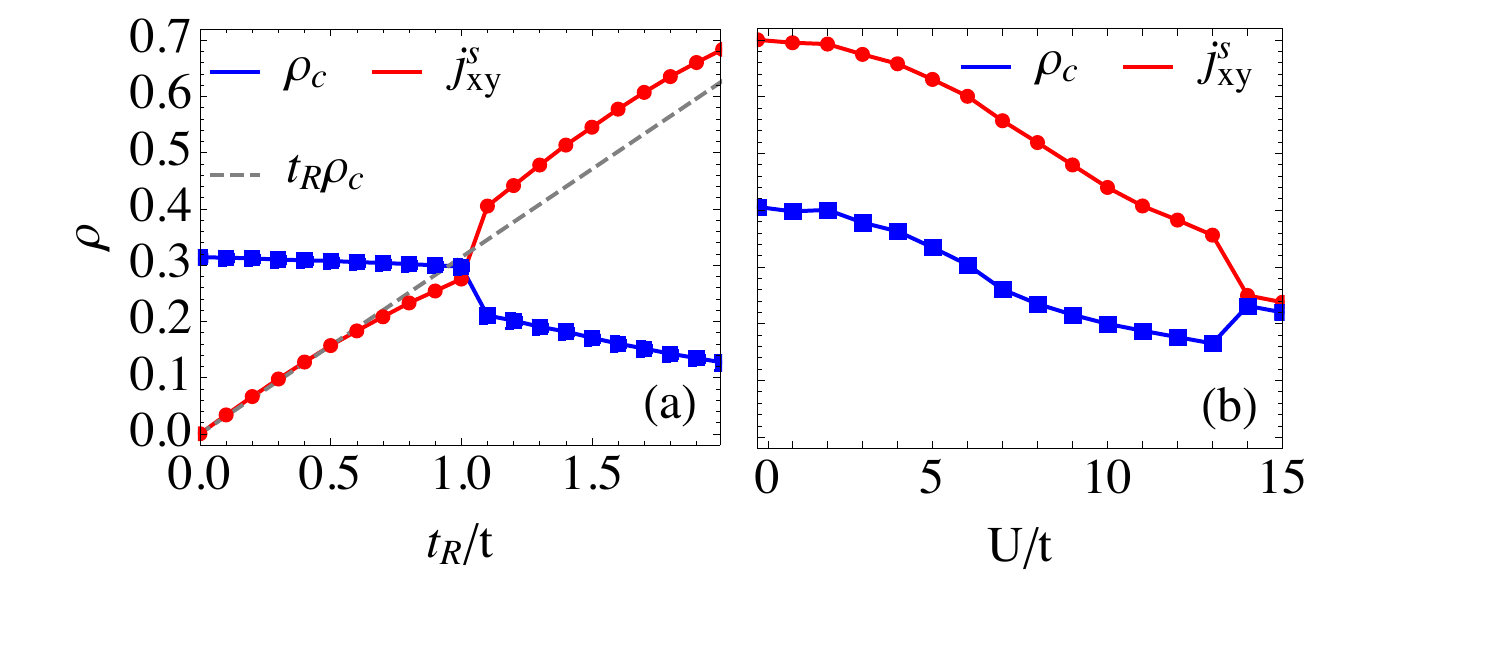}
\caption{(a) Bond-charge and spin-current as a function of $t_R/t$ for $U=15t$, the dashed line represents the product $\rho_c t_R$.   (b) Bond-charge and spin-current as a function of $U$ for $t_R=1.2t$}\label{fig-kin}
\end{center}
\end{figure}

It is interesting to note that the choice of squared clusters and the periodization scheme proposed in Eq. \eqref{eq:b}, preserving the symmetries of the original square lattice, correctly yield a vanishing spin-Hall current, {\sl i.e.}
\be
j_{s}^z=\int d\omega\sum_{\bf k} \rm{Tr}\lf[A(\omega,{\bf k})\sigma_z\rg] \sin(\bf k\, a_\mu) =0
\ee
with $A(\omega,{\bf k})$ denoting the spin-dependent spectral function $A_{\sigma\sigma'}(\omega,{\bf k})=-1/\pi{\rm Im}\,\lf[G_{\sigma\sigma'}(\omega,{\bf k})\rg]$.
The bond-charge, $\rho_c$,
$$\rho_c=\sum_{\bf k}\int \!\! A_0(\omega,{\bf k})\cos(k_x)d\omega,$$ and the in-plane equilibrium spin-current,\cite{rashba2003} $j^s_{xy}$, 
$$j^s_{xy}=-\frac{1}{2\pi}\sum_{\bf k}\int_{-\infty}^\infty \!\!{\rm Im}\lf[{\rm Tr}\lf[G(\omega,{\bf k} )\sin(k_x)\sigma_y\rg]\rg]d\omega,$$
have instead a non-trivial dependence on $U$ and $t_R$ as we show in Fig. \ref{fig-kin}(a-b).
We see that in the AF phase $j^s_{xy}$ grows linearly as a function of $t_R$  while the bond-charge is roughly constant,  for a certain value of $t_R$ corresponding to $j^s_{xy}=\rho_c$, the system undergoes the AF-HL transition.
 The transition yields a strong reduction of the bond charge and a strong increase of the spin-current, indeed due to Pauli principle standard tunneling is suppressed by ferromagnetic correlations while Rashba tunneling is enhanced.\\

\section{Rashba versus next-nearest neighbor tunneling}
\label{sec-tNNN}
 \begin{figure}[t!]
\begin{center}
\includegraphics[width=0.5\textwidth]{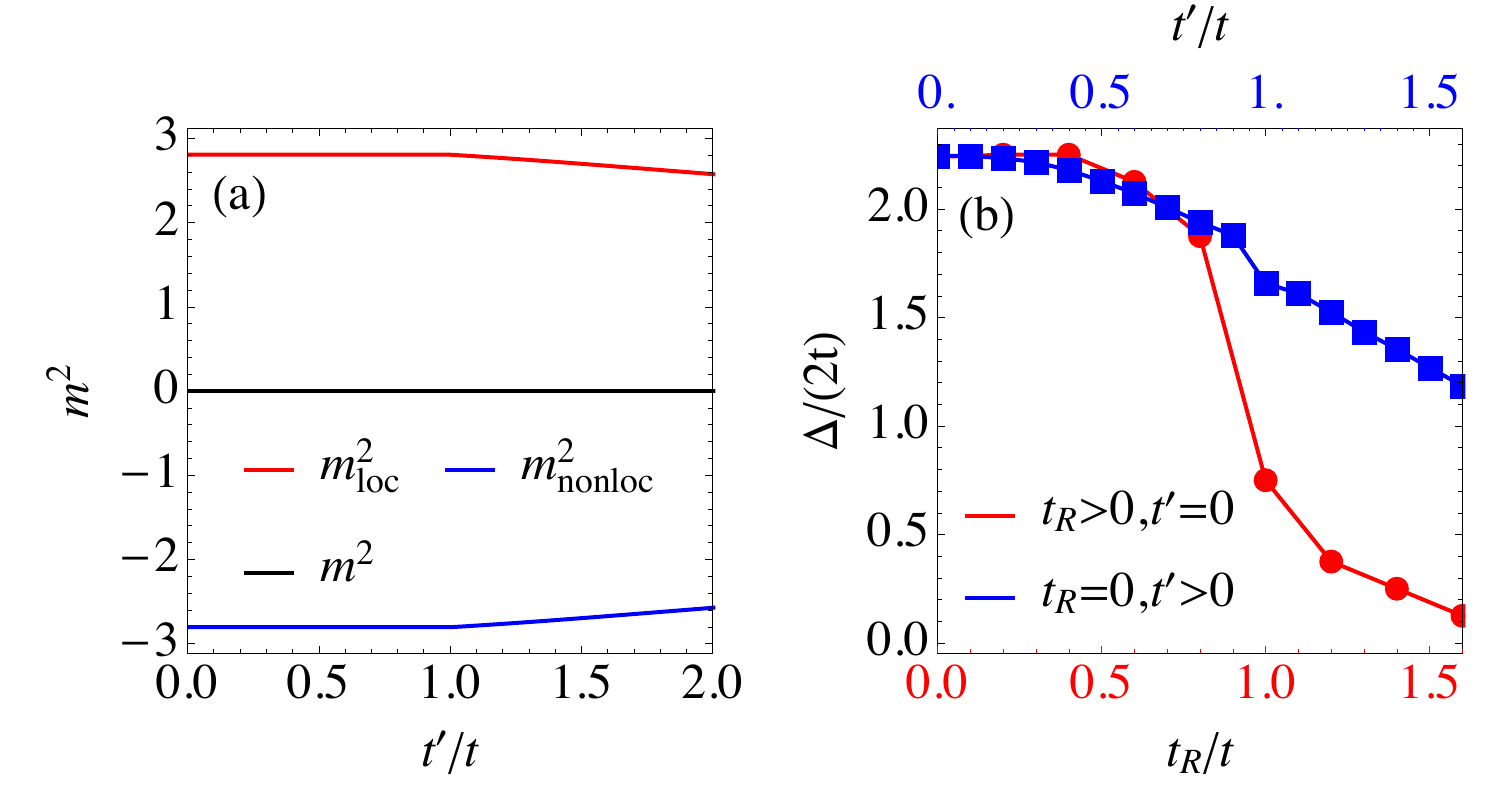}
\caption{(a) Local, non-local and total single-cluster magnetic moment  at $U=8t,\,t_R=0$ as a function of next-nearest-neighbor tunneling. (b) Spectral gap at $U=8t$ in the extended Hubbard model  and in the Rashba-Hubbard model. The value  $U=8t$ is chosen to have the transition in the Rashba-Hubbard model around $t_R\sim t$.  }\label{fig-gap-tprimo}
\end{center}
\end{figure} 
In this appendix, we elucidate the peculiarity of Rashba SOC as compared to next-nearest neighbor tunneling, focusing  in particular on the behavior of the charge gap.
We thus replace the RSOC term in Hamiltonian $H_0$,(Eq.\eqref{hopping+Rashba}) with a next-nearest-neighbor term defined as 
$$H_{NN}=-t'\sum_{\la\la ij\ra\ra}(c^\dag_{i\sigma}c_{j\sigma}+H.c.)$$
and we use CPT to calculate the spectral function and the charge gap.
At a mean field level,  Rashba  spin-orbit coupling and next-nearest neighbor tunneling, $t'$, have indeed similar consequences
on the density of states of electrons in a two-dimensional square lattice. Both these single-particle terms, break the perfect nesting of the Fermi surface shifting the van Hove singularity its usual location. They also have similar effects on the ground-state of a single plaquette.
Indeed, while for a standard Hubbard Hamiltonian,  at finite $U$ and $t$, the plaquette's ground state has $d$-symmetry, in the presence of  Rashba spin-orbit coupling\cite{brosco2018} or next-nearest neighbors tunneling,\cite{yao2010} it undergoes a transition to a state with $s$-symmetry. 

In spite of these similarities,   CPT yields very different spectral properties for the Rashba-Hubbard model and the  Hubbard model with next-nearest-neighbor tunneling. 

In particular, as shown in Figure \ref{fig-gap-tprimo}(a) $t^\prime$ has a weak effect on the magnetic moment of the plaquette and the non-local magnetization remain always negative indicating AF correlation.  By looking at the gap behavior, we can distinguish two regimes: in the first regime $t^\prime$ has roughly the same effect of $t_R$;
 this is the regime where  the ground state of the Rashba-Hubbard model is mostly a singlet. In the second regime, for $t^\prime>t$ the bandwidth of the extended Hubbard model starts to increase with increasing $t^\prime$, as one can see, however this  has a much weaker effect on the gap as compared to the screening induced by RSOC in the triplet ground-state.


\begin{thebibliography}{10}
\bibitem{FN2}More details on on the non-interacting band structure justifying the  expression of $t_{\rm eff}$  may be found in
appendix \ref{LSF}.
\end{thebibliography}
\end{document}